\documentclass[showkeys,showpacs,onecolumn]{revtex4}

\usepackage{epsfig}
\usepackage{amsfonts}
\usepackage{eufrak}
\usepackage {graphicx}
\usepackage {longtable}
\usepackage[T1]{fontenc}

\begin{document}

\title{Motion of a Particle with Isospin in the Presence of a Monopole}

\author{Rafael M. Fernandes}
\affiliation{Instituto de F\'\i sica "Gleb Wataghin", Universidade Estadual 
de Campinas, 13083-970, Campinas, SP, Brazil}
\email{rafaelmf@ifi.unicamp.br}

\author{Patricio S. Letelier}
\affiliation{Departamento de Matem\'atica Aplicada, Instituto de Matem\'atica, Estat\'\i stica e Computa\c{c}\~ao 
Cient\'\i fica, Universidade Estadual de Campinas, 13083-970, Campinas, SP, Brazil}
\email{letelier@ime.unicamp.br}

\begin{abstract} 
From a consistent expression for the quadriforce describing the interaction between a coloured
particle and gauge fields, we investigate the relativistic motion of a particle with isospin interacting with a BPS
monopole and with a Julia-Zee dyon. The analysis of such systems reveals the existence of unidimensional unbounded
motion and asymptotic trajectories restricted to conical surfaces, which resembles the equivalent case of
Electromagnetism. 
\end{abstract}



\maketitle

\section{Introduction}

Gauge theories were led to a major position in the foundations of Physics as the main tool to describe the Standard
Model of Elementary Particles. A great variety of work in this area has been done, and important achievements have
been reached by the study of non-Abelian symmetry groups such as $SU(2)$ or $SU(3)$. However, deeper analyses are
still necessary, specially in non-perturbative cases, in order to understand subtle aspects and to discover new
properties of these theories. Hence, an important tool to describe these systems is the semiclassical limit, with
which it is possible to obtain valuable insights about their quantum behaviour. And, in this context, an approach that
is largely used is to analyze the orbits of a single particle, which is the main objective of this article.

We start proposing an expression for the quadriforce that describes the interaction between a particle with isospin
and fields from gauge theories, which usually have scalar and vector components. Although such expression is well
established in the case of vector fields (the so called Wong's expressions \cite{wong}), it is still a matter of
investigation for scalar fields (\cite{feher}, \cite{azizi2}). To obtain an expression that takes into account both
contributions, we make use of Wong's equation combined with a gauge generalization of the quadriforce presented in our
previous work \cite{rafael}, which referred to scalar fields only.

With this formalism, we apply the resultant equations to two cases of physical significance: a coloured
particle interacting with a BPS monopole, which behaves asymptotically as a magnetic monopole, and with a Julia-Zee
dyon, which behaves asymptotically as a magnetic plus electric monopole. We observe that such systems present no
dynamical equilibrium points and have no apparent general analytical solutions. Hence, we particularize to simpler cases
in which the general properties of the motion of the particle can be well established: radial and asymptotic motions.

As far as the radial motion is concerned, we find that there are no bounded orbits for the particle: it always escapes
in a direction that is related to the sign of the radial component of its isospin. Analyzing the asymptotic motion, in
the limits of low (Newtonian) and high (ultra-relativistic) velocities, we observe the existence of bounded orbits
under certain conditions. The remarkable property, however, is that all of the asymptotic trajectories are constrained
to a conical surface in both limits. And this is exactly the case of Classical Electromagnetism, in which the motion
of a point charge interacting with a magnetic monopole is restricted to the surface of a cone.

This article is divided as follows: in Section \ref{sec:BPS} we derive the equations of motion for the particle in the
presence of the BPS monopole and analyze it globally; in Section \ref{sec:radial}, we specify them to radial motion
and, in Section \ref{sec:asympt}, to asymptotic motion. The case in which the monopole involved is the Julia-Zee dyon
is presented in Section \ref{sec:dyon} and conclusions are described in Section \ref{sec:concl}.

\section{Coloured Particle in the Presence of a BPS Monopole} \label{sec:BPS}

Firstly, we derive the general equations that describe the interaction of a particle with isospin and gauge fields.
Two sets of equations are requested: the one referring to the quadriforce and the one to the isospin's precession. As
we have previously explained, we propose an expression for the quadriforce that is a combination of Wong's equation
and a generalization of the equation proposed in our last article. This was originally conceived to take into
account scalar fields only:

\begin{equation} \label{original}
m \frac{d^2 x^{\mu}}{d \tau^2}  =  g \left( \eta^{\mu \nu} - \frac{d x^{\mu}}{d \tau} \frac{d x^{\nu}}{d \tau}
\right) \partial_{\nu} \phi^{a} I^a
\end{equation}

In the above expression, $x^{\mu}$ are the coordinates of the particle, $\tau$ is its proper time, $I^a$ are the
components of its isospin, $g$ is the coupling constant, $\eta^{\mu \nu}$ is the Minkowskian metric and $\phi^a$ are
the scalar fields. It is straightforward to generalize (\ref{original}) to gauge fields: we only have to replace the
ordinary derivatives by covariant ones:

\begin{equation} 
m \frac{d^2 x^{\mu}}{d \tau^2}  =  g \left( \eta^{\mu \nu} - \frac{d x^{\mu}}{d \tau} \frac{d x^{\nu}}{d \tau}
\right) D_{\nu} \phi^{a} I^a
\end{equation}

To include the self-interaction of the gauge fields as well, it is necessary to introduce Wong's expression,
which leads us to:

\begin{equation} \label{quadriforce}
m \frac{d^2 x^{\mu}}{d \tau^2}  =  g \left( \eta^{\mu \nu} - \frac{d x^{\mu}}{d \tau} \frac{d x^{\nu}}{d \tau}
\right) D_{\nu} \phi^{a} I^a + g F^{a \mu}_{\nu} \frac{d x^{\nu}}{d \tau} I^a
\end{equation}

\noindent where $F^{a \mu}_{\nu}$ are the components of the strength tensor of the fields. 

The isospin's precession equation is obtained just in the same way: we make use of the expression proposed in our last
work generalized to gauge fields (which recovers Wong's expression in the case of vector fields only):

\begin{equation} \label{isospin}
\frac{\partial I^a}{\partial \tau}  +  g \epsilon^{abc} \left( D_{\mu} \phi^{b} + A^{b}_{\mu} \right) I^c \frac{d
x^{\mu}}{d \tau} = 0
\end{equation}

Therefore, equations (\ref{quadriforce}) and (\ref{isospin}) are the equations of motion of a coloured particle 
interacting with gauge fields. 

The BPS monopole is described by the fields (\cite{bogo}, \cite{prasad}):

\begin{equation}\label{BPS}
\phi^a  =  x^a \frac{H(r)}{gr^2} \, \, \, \, \, ; \, \, \, \, \,
A_{a}^{i} = \epsilon_{a i j} x^j \frac{1-K(r)}{gr^2}
\end{equation}

\noindent where:

\begin{equation}\label{BPS2}
H(r)  = \frac{gFr}{\tanh{gFr}} - 1 \, \, \, \, \, ; \, \, \, \, \,
K(r)  = \frac{gFr}{\sinh{gFr}}
\end{equation}

\noindent and $F$ is a constant. Hence, by applying them to the equations (\ref{quadriforce}) and (\ref{isospin}), we 
get the following system:   

\begin{eqnarray} \label{system}
m \vec{a} & = & \frac{HK}{\gamma^2 r^2} \vec{I} + \frac{rH' - H(1+K)}{\gamma^2 r^4} \left( \vec{I} \cdot \vec{r}
\right) \vec{r} 
+\frac{K^2 - rK'-1}{\gamma r^4} \left( \vec{I} \cdot \vec{r} \right) \left( \vec{v} \times \vec{r}
\right) + \nonumber \\
& + & \frac{K'}{\gamma r} \left( \vec{v} \times \vec{I} \right) \\
\dot{\vec{I}} & = & \frac{1-K}{r^2} \left[ \vec{I} \times \left( \vec{v} \times \vec{r} \right) \right] +
\frac{HK}{r^2} \left( \vec{I} \times \vec{v} \right) + \frac{rH'-H (1+K)}{r^2} \left( \vec{r} \cdot \vec{v} \right) 
\left( \vec{I} \times \vec{r} \right) \nonumber
\end{eqnarray}

We can write it as a dynamical system and look for equilibrium points to study its stability. Imposing the 
equilibrium condition and using a base of co-moving coordinates,

\begin{equation}\label{base}
\left\{ \hat{r} , \hat{w} = \frac{\vec{r} \times \vec{v}}{\mid \vec{r} \times \vec{v} \mid}, \hat{u} = \frac{\vec{r}
\times \vec{w}}{\mid \vec{r} \times \vec{w} \mid} \right\}
\end{equation}

\noindent it is straightforward to conclude that there are no equilibrium points. Therefore, we particularize
(\ref{system}) to more restricted conditions and study the general properties of the resultant orbits.

\subsection{Radial Motion} \label{sec:radial}

If the particle is initially moving in the radial direction, $\vec{v}_{0} \parallel \vec{r}$, and its isospin is also 
in the same direction, $\vec{I}_{0} \parallel \vec{r}$, we can use equations (\ref{system}) to show that the 
particle will remain in the radial direction according to:

\begin{eqnarray} \label{radial}
m\vec{a} & = & - \left( 1 - \dot{r}^2 \right) \frac{dV(r)}{dr} \hat{r} \nonumber \\
\vec{I} & = & I_{0} \hat{r} = \alpha \hat{r} 
\end{eqnarray}

\noindent where the function $V(r)$ is defined as:

\begin{equation}\label{potencial}
V(r) = - \alpha \frac{H(r)}{r} = \alpha \left[ - \frac{gF}{\tanh{gFr}} + \frac{1}{r} \right]
\end{equation}

To study this unidimensional motion, we can firstly analyze it in the Newtonian limit, $\mid \dot{r} \mid \ll 1$. It
is clear that, in this case, the function $V(r)$ is nothing but the potential energy of the particle due to its
interaction with the monopole; moreover, this function has the following properties:

\[
\left\{  \begin{array}{ccc}
\displaystyle{\lim_{r \rightarrow 0}} V(r) &  = & 0 \\
\displaystyle{\lim_{r \rightarrow \pm \infty}} V(r) & = & \mp \alpha g F  \\
\frac{dV}{dr} & \neq & 0
\end{array} \right. \Rightarrow \; \left\{ \begin{array}{ll}
V(r) < 0 & \; \; \mbox{for all} \; \; r>0 \\
V(r) > 0 & \; \; \mbox{for all} \; \; r<0
\end{array} \right.
\]

Such characteristics of the potential energy can also be seen in figure 1, in which the graphic of the function is
shown for the case $g=F=1$. Hence, we conclude that there are no bounded orbits in this limit, and the particle will
always escape. Of course, in order for the particle not to leave the region of validity of the Newtonian limit, it is
necessary to guarantee that the constants involved satisfy the condition $E \ll m/2 + \mid \alpha \mid g F$.

\begin{figure*}
\centering
\epsfig{width=6cm, height=6cm, file=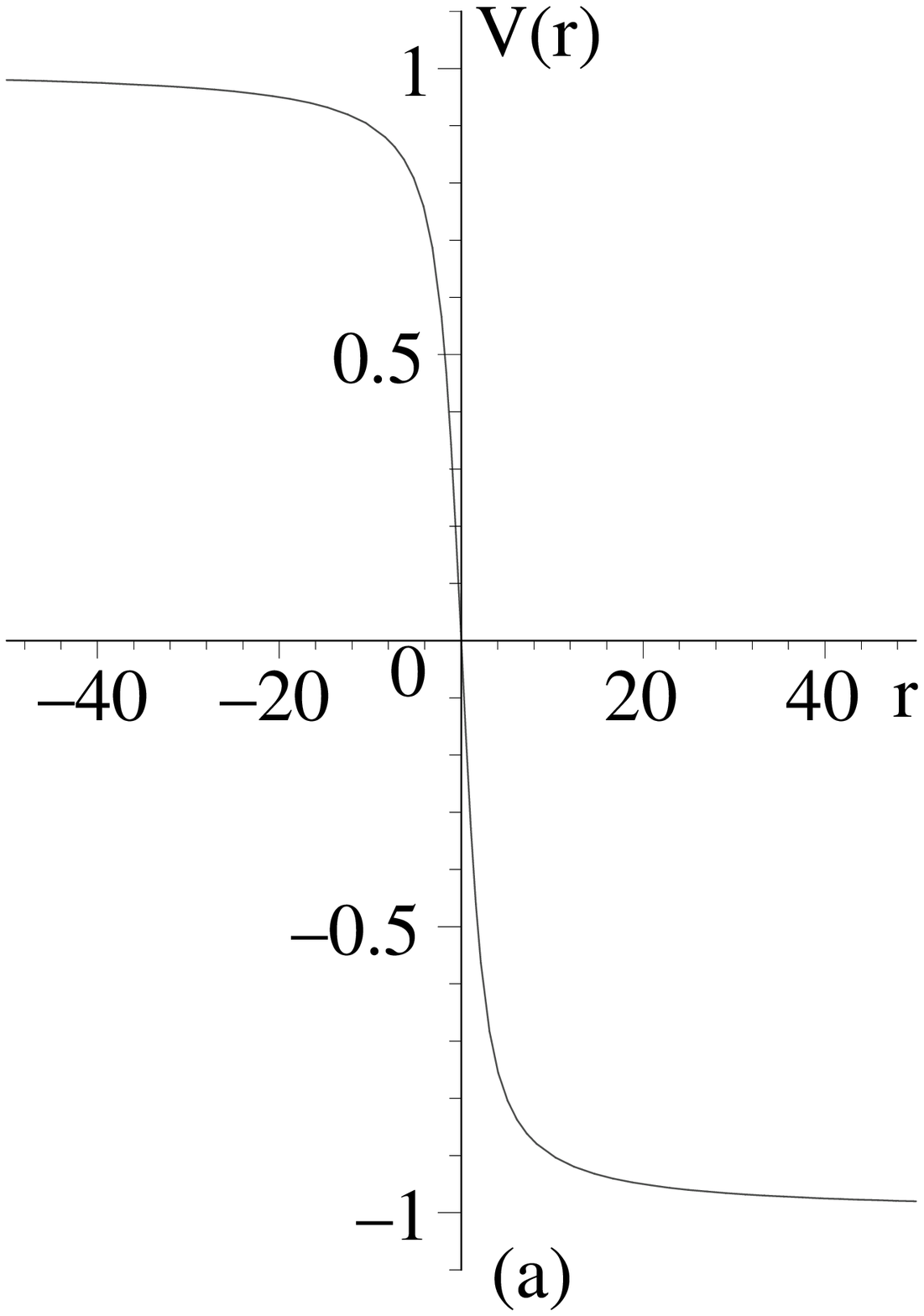} \hspace{1cm}
\epsfig{width=6cm, height=6cm, file=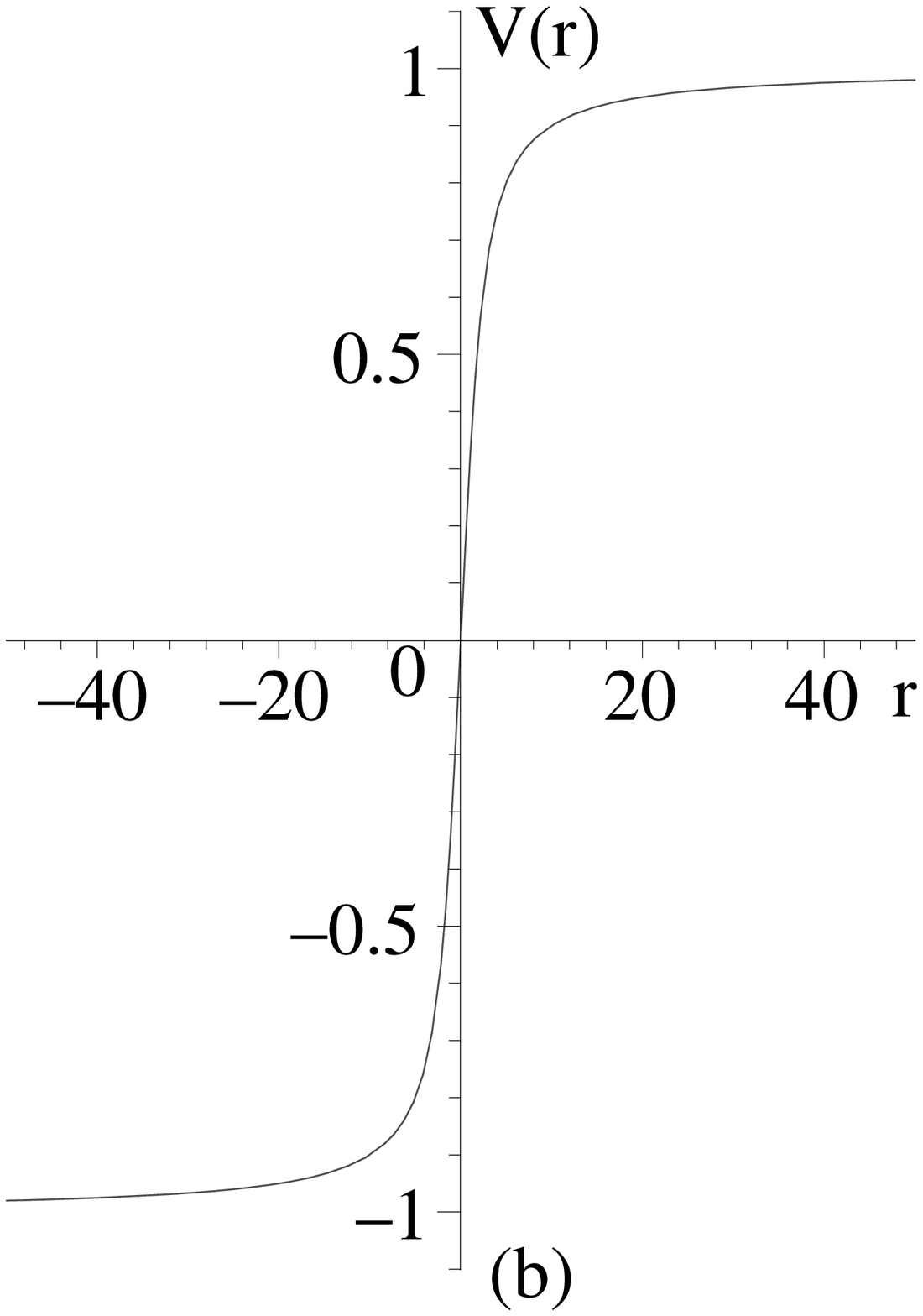}
\caption{Potential energy of a coloured particle in the presence of a BPS field with $g=F=1$, in the Newtonian limit 
and in radial motion with:(a) $\alpha=1$, (b) $\alpha=-1$.}
\label{potencial:BPS}
\end{figure*}

Now, we can analyze equations (\ref{radial}) in the relativistic case. Although an analytical solution is not 
available, we can infer the main properties of the resultant motion by seeking a constant of motion. Taking $T=m 
\gamma$, the total relativistic kinetic energy of the body, we obtain from (\ref{radial}) the following constant of 
motion:

\begin{equation} \label{energy}
E = m \ln{\frac{T}{m}} + V(r)
\end{equation}

This can be interpreted as the total relativistic energy of the particle, since that, in the Newtonian limit, the
first term reduces to the well known expression of the classical kinetic energy. Analyzing this constant, it is clear
that there is no more than one return point (i.e., points in which $\dot{r}=0$) in the system, implying that no
bounded orbits are allowed, and again the particle always escapes. The equation describing such motion is given
implicitly by:

\begin{equation}
\int_{r_0}^{r} \frac{dr'}{\sqrt{1 - \frac{m^2}{\kappa^2} \mbox{e}^{2V(r')/m}}} = \pm t
\end{equation}
                                                                                                                             
\noindent where:
                                                                                                                             
\[
\kappa = T(r_0) \mbox{e}^{V(r_0)/m}
\]

\subsection{Asymptotic Motion} \label{sec:asympt}

An interesting feature of equations (\ref{system}) is their behaviour when $r \gg 1$, since, in this limit, the BPS 
solution behaves as a magnetic monopole. Applying this condition, we are led to the new system:

\begin{eqnarray}\label{asympt}
m\vec{a} & = & \frac{\alpha}{\gamma^2 r^3} \vec{r} + \frac{\alpha}{\gamma r^3} \left( \vec{r} \times \vec{v}
\right) \nonumber \\
\dot{\vec{I}} & = & \frac{\left( \vec{r} \cdot \vec{v} \right)}{r^2} \left( \vec{I} \times \vec{r} \right)
\end{eqnarray}

\noindent in which, again, $\alpha$ is the radial component of the particle's isospin. These equations can indeed be 
simplified by changing the coordinate system to that of (\ref{base}); hence, writing the isospin vector as:

\begin{equation}
\vec{I} = \alpha \hat{r} + \beta \hat{w} + \rho \hat{u}
\end{equation}

\noindent we obtain that the second equation of (\ref{asympt}) has the following solution:

\begin{eqnarray}
\alpha(t) & = & \alpha(0) \nonumber \\
\beta(t) & = & \beta(0) \cos{\Delta r(t)} + \rho(0) \sin{\Delta r(t)} \nonumber \\
\rho(t) & = & \rho(0) \cos{\Delta r(t)} - \beta(0) \sin{\Delta r(t)}
\end{eqnarray}

\noindent where $\Delta r(t) = r(t) - r(0)$. Thus, the motion of the particle in the external (Minkowskian) 
space is not coupled to the motion in the internal (isospin) space, where its isospin precessionates according to the 
previous expressions. So, we have to solve the first equation of (\ref{asympt}) to completely describe the particle's 
movement. However, we seem to have neither an analytical solution nor enough constants of motion for the equation. 
Hence, we make further approximations and study the system in two very distinct limits: the Newtonian limit ($v \ll 1$) 
and the ultra-relativistic limit ($v \approx 1$).

In the Newtonian limit, the equation is reduced to:

\begin{equation}\label{assintnewton}
m\vec{a}  = \frac{\alpha}{r^3} \vec{r} + \frac{\alpha}{r^3} \left( \vec{r} \times \vec{v} \right)
\end{equation}
     
We observe that (\ref{assintnewton}) is very similar to the Lorentz equation of Electromagnetism corresponding to a 
particle with charge $\alpha$ interacting with an electric field generated by a unit charge and a magnetic field 
generated by a unit magnetic monopole. The first term is due to the scalar fields and, the second, due to the vector 
fields. Equation (\ref{assintnewton}) has three constants of motion that can be directly obtained:

\begin{equation}
E = \frac{mv^2}{2} + \frac{\alpha}{r} \, \, \, \, \, ; \, \, \, \, \,
\vec{J} = m \left( \vec{r} \times \vec{v} \right) + \alpha \hat{r} \, \, \, \, \, ; \, \, \, \, \,
L = |m \left( \vec{r} \times \vec{v} \right)|
\end{equation}

It is clear that the first one is the total energy of the particle, the second is its total angular momentum (see, for
instance \cite{momang}) and the third, the module of its orbital angular momentum. To determine the equations of
motion, we choose spherical coordinates and make the $z$ axis coincide with $\vec{J}$. Therefore, it is
straightforward that the particle is restricted to move on the surface of a cone whose axis is in the same direction
of $\vec{J}$ and whose half-angle of opening is given by:

\begin{equation}
\cos{\theta} = \frac{\alpha}{J}
\end{equation}

In addition, combining the two other remaining constants, we obtain:

\begin{equation}\label{Vefetivo}
E = \frac{m\dot{r}^2}{2} + \frac{\alpha}{r} + \frac{L^2}{2mr^2} = \frac{m\dot{r}^2}{2} + V_{eff}(r)
\end{equation}

It is now possible to analyze the motion in the radial direction by this single constant. The shape of the effective 
potential is shown in figure 2, considering the two possible signs of $\alpha$. We can conclude that, if $\alpha > 0$, 
the particle's energy has to be positive and there is only one return point in:

\begin{equation} \label{retorno1}
\bar{r} = \frac{\alpha}{2E} \left( 1 + \sqrt{1 + \frac{2L^2 E}{m \alpha ^2}}  \right)
\end{equation}

\begin{figure*}
\centering
\epsfig{width=6cm, height=6cm, file=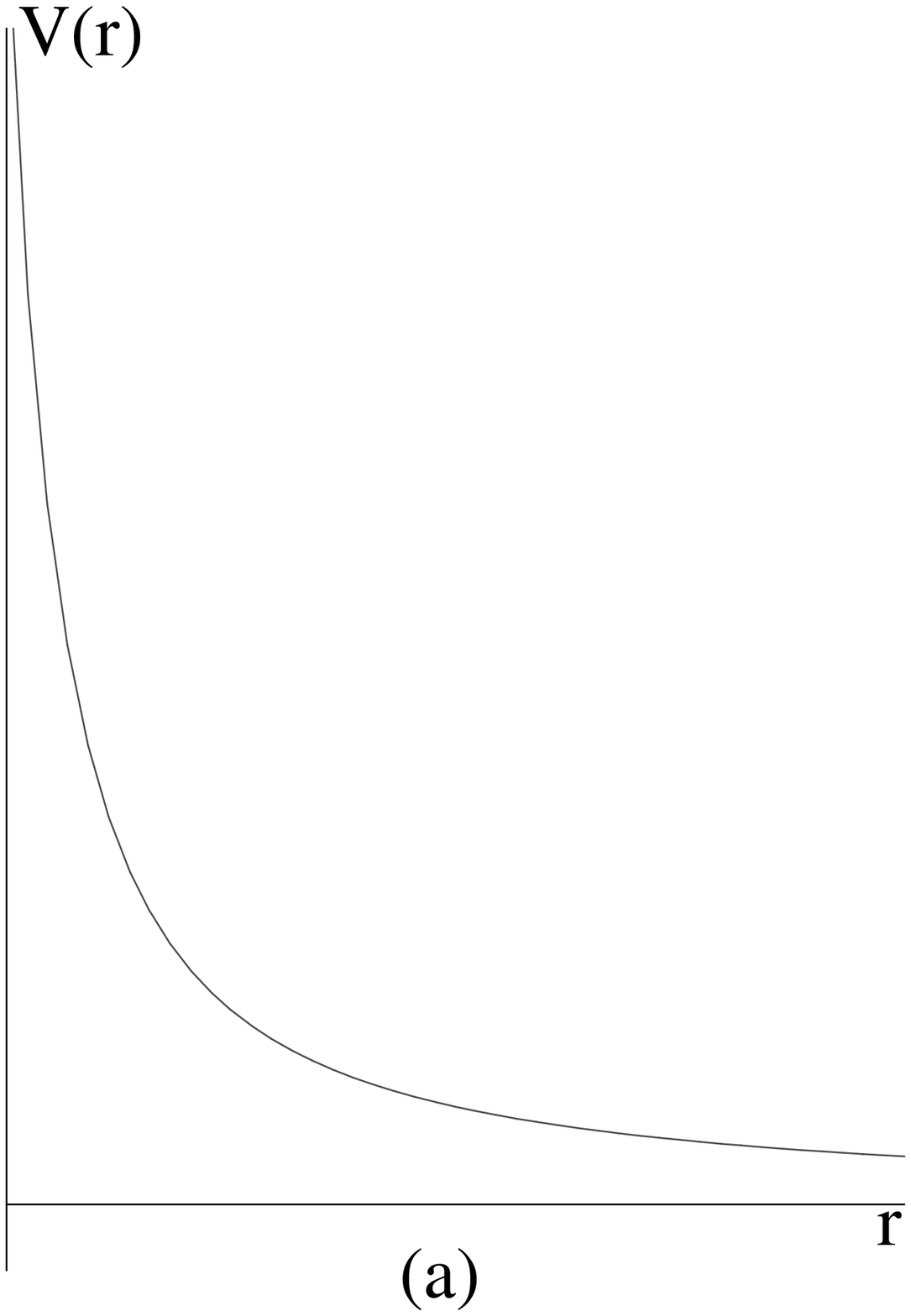} \hspace{1cm}
\epsfig{width=6cm, height=6cm, file=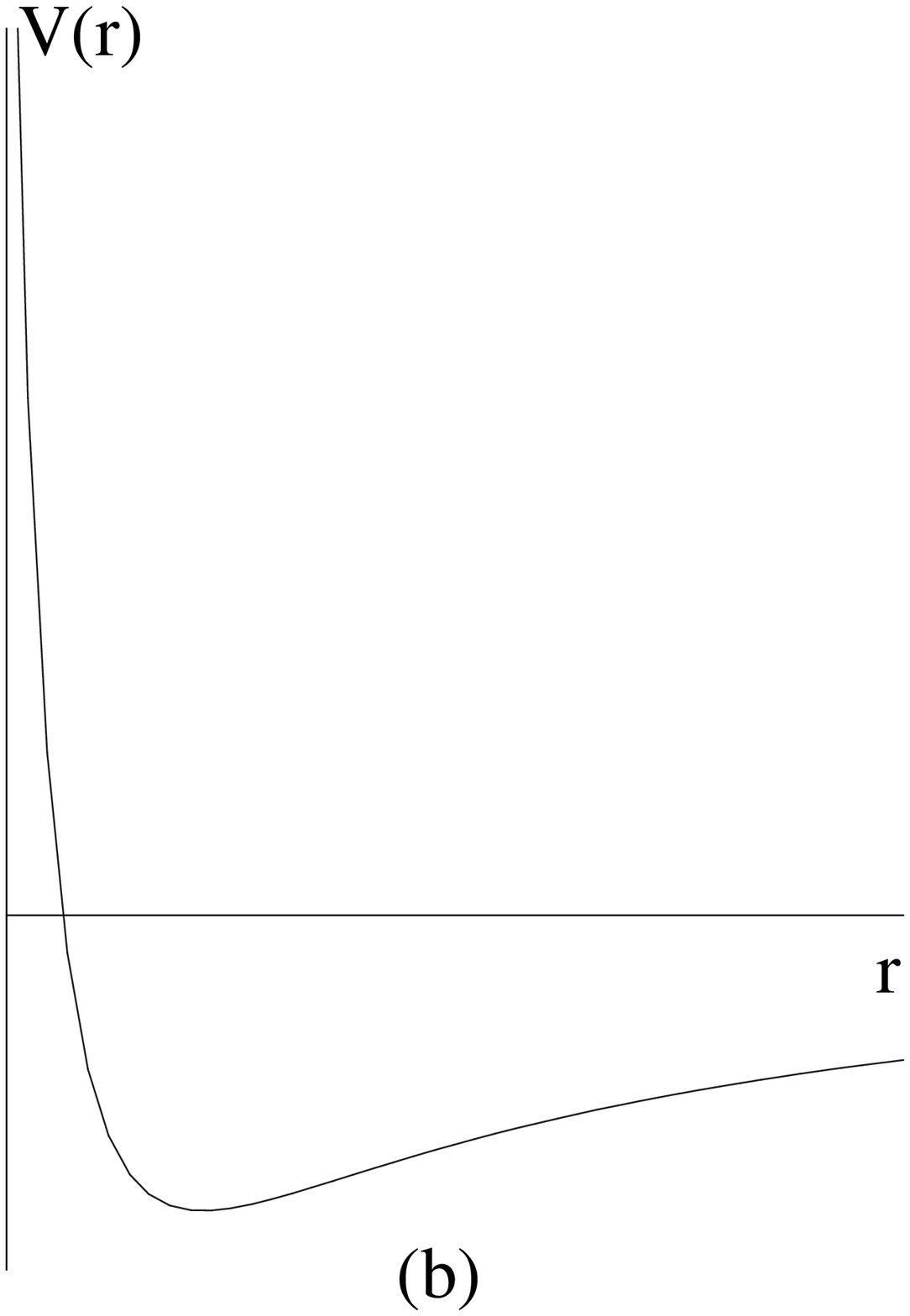}
\caption{Shape of the effective radial potential for (a) $\alpha>0$ and (b) $\alpha<0$.}
\label{fig:vefetivo}
\end{figure*}

It is important to notice that this "return point" is actually a point in which $\dot{r}$ vanishes, and not the
velocity $v$. Therefore, the particle will also escape. Moreover, it is necessary that $E \ll \alpha$ and $E \ll m/2$
for the return point to be in the asymptotic region and for the particle to remain in the Newtonian limit,
respectively.

However, in the case in which $\alpha < 0$, the particle is allowed to have positive or negative energy. If it is 
positive, there is again only one return point given by:

\begin{equation} \label{retorno2}
\bar{r} = - \frac{\alpha}{2E} \left(\sqrt{1 + \frac{2L^2 E}{m \alpha ^2}} - 1  \right)
\end{equation}

\noindent meaning that there are no bounded orbits. But if the energy takes negative values, there are two return 
points, given by:

\begin{equation} \label{retorno3}
\bar{r}_{\pm} = -\frac{\alpha}{2|E|} \left(1 \pm \sqrt{1 - \frac{2L^2 |E|}{m \alpha ^2}}  \right)
\end{equation}

\noindent which means that the movement is bounded. In addition, there is one equilibrium point, in which the particle 
just precessionates along the intersection between the cone and the sphere of radius given by: 
                                                                               
\begin{equation} \label{pontoequil}
r_{eq} = - \frac{L^2}{m \alpha}
\end{equation}

Of course, some conditions regarding all these constants have to be satisfied in order to guarantee that these return
and equilibrium points are in the asymptotic region and that the particle remains with low velocities. These are:  
$|E|\ll |\alpha|$, $L^2 \gg m |\alpha|$, $L^2 |E| < m \alpha^{2}/2$ and $|E| \ll m/2$.

The analytical expression of the radial coordinate, $r(t)$, can be obtained directly by solving the differential
equation (\ref{Vefetivo}). We note that, if $E>0$, the solution is:

\begin{eqnarray}\label{sol:edonewton}
& & \left[ \frac{m}{2E}\sqrt{\frac{2E}{m}r^2 - \frac{2 \alpha}{m}r - \frac{L^2}{m^2}} + \frac{\alpha 
\sqrt{m}}{(2E)^{3/2}}  \right. \nonumber \\ 
& & \left. \ln \left( \frac{ \sqrt{8E} \sqrt{2Er^2 - 2 \alpha r - L^2/m} + 4Er - 2 \alpha}{m} \right) \right] 
^{r(t)}_{r_0}= \pm t  
\end{eqnarray}

\noindent while, for $E<0$, it is given by:

\begin{eqnarray}\label{sol:edonewton2}
& & \left[ \frac{m}{2E}\sqrt{\frac{2E}{m}r^2 - \frac{2 \alpha}{m}r - \frac{L^2}{m^2}} - \right. \nonumber \\ 
& & \left. - \frac{\alpha \sqrt{m}}{(2|E|)^{3/2}}\arcsin\left(\frac{4Er -2 \alpha}{\sqrt{4 \alpha^2 + 
8EL^2/m}}\right) \right] ^{r(t)}_{r_0}= \pm t
\end{eqnarray}

The sign in the equations has to be chosen according to the initial radial velocity of the particle and changes 
every time it crosses one of the return points presented. 

To complete the description of the movement, we have to determine the expression for the azimuthal angle, $\phi(t)$.
Writing the total velocity in spherical coordinates, $v^2 = \dot{r}^2 + r^2 \dot{\theta}^2 + r^2 \sin^2{\theta}
\dot{\phi}^2$, and using the constants of motion, we arrive at the implicit equation:

\begin{equation}\label{phi}
\phi(t) = \phi(0) \pm \frac{J}{m} \int_{0}^{t} \frac{dt'}{r^2(t')}
\end{equation}

In the case where the particle precessionates along the intersection between the cone and the sphere of radius
(\ref{pontoequil}), this last equation can be explicitly solved, and the result is given by:

\begin{equation}
\phi(t) = \phi(0) \pm \frac{mJ\alpha ^2}{L^4} t^2
\end{equation}

Now that we have analyzed the motion in the Newtonian limit, we can move to the opposite limit, the ultra-relativistic 
motion. In this case, the equation of motion is reduced to:

\begin{equation}\label{ultrarel}
m \vec{a} = \frac{\alpha}{\gamma r^3} \left( \vec{r} \times \vec{v} \right)
\end{equation}
                               
It is clear that the velocity of the particle is a constant of motion, implying that once the body is in the 
ultra-relativistic limit, it will remain there. Moreover, two other constants can be readily obtained:

\begin{equation}
\vec{J}_{rel} = \frac{m \left( \vec{r} \times \vec{v} \right)}{\sqrt{1 - v^2}} + \alpha \hat{r} \, \, \, \, \, ; \, \, 
\, \, \,
L_{rel} =  \frac{|m \left( \vec{r} \times \vec{v} \right)|}{\sqrt{1 - v^2}}
\end{equation}

They are, respectively, the total relativistic angular momentum and the module of the orbital relativistic angular
momentum. Hence, the orbits of the particle are again confined to a conical surface whose axis coincides with
$\vec{J}_{rel}$ and whose half-angle of opening is:

\begin{equation}\label{conerel}
\cos{\theta_{rel}} = \frac{\alpha}{J_{rel}}
\end{equation}

By using the fact that the module of the orbital angular momentum and the velocity are constants, it is possible to
get a differential equation for $r(t)$ whose solution is given explicitly by:

\begin{equation} \label{r_rel}
r(t) = r_0 \sqrt{\sin^2{\omega_0} + \left( |\cos{\omega_0}| \pm \frac{vt}{r_0} \right)^2 }
\end{equation}

\noindent where $\omega_0$ is the initial angle between the velocity and the position of the particle and the sign 
of the equation has to be chosen according to the initial sign of the radial velocity of the particle. We can see that, 
even if the particle is initially going in the direction of the origin, it will never reach this point, but will stop 
at $r=r_0 |\sin{\omega_0}|$ and return, escaping to infinity. Therefore, in this limit of high velocities, there is no 
bounded motion. 

To complete the description of the particle's movement, it is straightforward to obtain the azimuthal coordinate
explicitly by using the constants of motion and the equation for $r(t)$:

\begin{equation} \label{phi_rel}
\phi(t) =\phi(0) \pm \frac{J}{L} \left[ \arctan{\left( \frac{|\cos{\omega_0} | \pm vt/ r_0}{
\sin{\omega_0}} \right)} - |\omega_0 - \frac{\pi}{2}| \right]
\end{equation}  

\section{Coloured Particle in the Presence of a Julia-Zee Dyon} \label{sec:dyon}

Now, we can apply the equations describing the interaction between a coloured particle and gauge fields, 
(\ref{quadriforce}) and (\ref{isospin}), to the case in which the fields are those from the Julia-Zee dyon solution, 
that are given by \cite{juliazee}:

\begin{equation}\label{dyon}
\phi^a = x^a \frac{\overline{H}(r)}{gr^2} \, \, \, \, \, ; \, \, \, \, \,
A_{a}^{i} = \epsilon_{a i j} x^j \frac{1-K(r)}{gr^2} \, \, \, \, \, ; \, \, \, \, \,
A_{a}^{0} = x^a \frac{J(r)}{gr^2}
\end{equation}
                                                                                                                             
\noindent in which the new functions are:
                                                                                                                             
\begin{equation}\label{dyon2}
\overline{H}(r) = H(r) \cosh{\lambda} \, \, \, \, \, ; \, 
\, \, \, \,
J(r) =  H(r) \sinh{\lambda}
\end{equation}

\noindent and $\lambda$ is a constant. We can see that this solution has plenty of similarities with the BPS
monopole. The main difference is that the dyon vector field has a non-vanishing time component, due to the existence
of the non-vanishing constant $\lambda$. It is this component that asymptotically generates an electric field that
is absent in the BPS monopole solution, inviting us to interpret the dyon as an entity with magnetic plus electric
charge.  Replacing equations (\ref{dyon}) and (\ref{dyon2}) in (\ref{quadriforce}) and (\ref{isospin}), we reach the
following equations of motion for the system:

\begin{eqnarray}
m \vec{a} & = & \frac{HK}{\gamma^2 r^2} \left[ \vec{I} \left( \cosh{\lambda} + \gamma \sinh{\lambda} \right) -
\gamma \sinh{\lambda} \left( \vec{I} \cdot \vec{v} \right) \vec{v} \right] + \nonumber \\
& & + \frac{rH' - H(1+K)}{\gamma^2 r^4} \left( \vec{I} \cdot \vec{r} \right) \cdot  \left[ \left( \cosh{\lambda} +
\gamma \sinh{\lambda} \right) \vec{r} - \gamma \sinh{\lambda} \left( \vec{r} \cdot \vec{v} \right) \vec{v} \right] + 
\nonumber \\
& & + \frac{K^2 - rK'-1}{\gamma r^4} \left( \vec{I} \cdot \vec{r} \right) \left( \vec{v} \times \vec{r} \right) + 
\frac{K'}{\gamma r} \left( \vec{v} \times \vec{I} \right) \label{movdyon} \\
\dot{\vec{I}} & = & \frac{1-K}{r^2} \left[ \vec{I} \times \left( \vec{v} \times \vec{r} \right) \right] +
\frac{HK\cosh{\lambda}}{r^2} \left( \vec{I} \times \vec{v} \right) + \nonumber \\
& & + \frac{\left[rH'-H (1 + K - \tanh{\lambda})\right] }{r^2} \cdot \cosh{\lambda} \left( \vec{r} \cdot \vec{v} 
\right) \left( \vec{I} \times \vec{r} \right) 
\label{movdyon2}
\end{eqnarray}

\noindent which bears some resemblance with the equations of motion when the interacting field is the BPS monopole. It
is straightforward to note that there are no equilibrium points for this system either. Thus, as an analytical
solution does not seem available, we study special cases for this system, as we have done for the case of the BPS
monopole. First, one can note that the particle can develop radial motion if its initial isospin and its initial
velocity point both to the radial direction; the equations of motion resultant from these conditions are:

\begin{eqnarray} \label{radial_dyon}
m\vec{a} & = & - \frac{1}{\gamma ^2} \frac{dV(r)}{dr} \left( \cosh{\lambda} + \frac{\sinh{\lambda}}{\gamma} \right) 
\hat{r} \nonumber \\
\vec{I} & = & I_{0} \hat{r} = \alpha \hat{r}
\end{eqnarray}

\noindent where $V(r)$ is still the function (\ref{potencial}). We can also see that, in the Newtonian limit, the 
equations of motion are the same as the ones referring to the BPS monopole, as long as we replace the constant $\alpha$ 
by an "effective constant" \, $\overline{\alpha} = \alpha \cosh{\lambda}$ that takes into account the electric charge 
carried by the dyon. Hence, the qualitative description of the motion is the same: no bounded orbits are allowed and 
the particle escapes to infinity.

In the relativistic case, we can study the expressions in (\ref{radial_dyon}) by searching for constants of motion.  
As in the BPS monopole case, a constant that can be interpreted as the total relativistic energy of the
particle can be found and is given by:

\begin{equation}
E = \frac{m}{\cosh{\lambda}}\ln{\left(T/m + \tanh{\lambda}\right)} + V(r)
\end{equation}

\noindent where $T=m \gamma$, again. When we analyze the conditions in which return points occur, it is possible to
conclude that, as in the BPS monopole case, there is at most one return point. This means that no bounded orbits are
allowed even in the relativistic case. Therefore, the global behaviour of the radial orbits of systems (\ref{radial})
and (\ref{radial_dyon}) are very similar.

Another special case of equations (\ref{movdyon}) and (\ref{movdyon2}) is the asymptotic limit. Making $r \gg 1$, we
are led to the following expressions:

\begin{eqnarray}\label{assintdyon}
m\vec{a} & = & \frac{\alpha}{\gamma^2 r^3} \left[ \left( \cosh{\lambda} + \gamma \sinh{\lambda} \right) \vec{r} -
\gamma \sinh{\lambda} \left( \vec{r} \cdot \vec{v} \right) \vec{v} \right] + + \frac{\alpha}{\gamma r^3} 
\left( \vec{r} \times \vec{v} \right)  \nonumber \\
\alpha (t) & = & \alpha (0) \nonumber \\
\beta(t) & = & \beta(0) \cos{\left( \Delta r(t) \mbox{e}^{-\lambda} \right)} + \rho(0) \sin{\left( \Delta r(t)
\mbox{e}^{-\lambda} \right)} \nonumber \\
\rho(t) & = & \rho(0) \cos{\left( \Delta r(t) \mbox{e}^{-\lambda} \right)} - \beta(0) \sin{\left( \Delta r(t)
\mbox{e}^{-\lambda} \right)}
\end{eqnarray}

\noindent where

\[
\vec{I} = \alpha \hat{r} + \beta \hat{w} + \rho \hat{u}
\]

Once more, as in the case of the BPS monopole, the motion in the external space is not coupled to the motion in the 
internal space, where the isospin precessionates. The only difference is in the frequency of the oscillations of the 
isospin vector, changed by the factor $\mbox{e} ^{- \lambda}$ due to the dyon's electric charge. 

Now, we analyze the equation of motion in the external space in the Newtonian limit, in which they are reduced to:

\begin{equation}\label{assintnewtondyon}
m\vec{a}  = \frac{\alpha \mbox{e}^{\lambda}}{r^3} \vec{r} + \frac{\alpha}{r^3} \left( \vec{r} \times \vec{v} \right)
\end{equation}

This expression has the same similarity to the Lorentz equation in Eletromagnetism as described for equation
(\ref{assintnewton}). The difference now is the presence of the factor $\mbox{e}^{\lambda}$ in the term referring to
the electric field, due to the contribution of the dyon's electric charge. This means that the particle is now
interacting with an electric charge $\mbox{e}^{\lambda}$, and not with an unit one, as it was the case for the BPS 
monopole.

Three constants of motion can be found again, and are given by:

\begin{equation}
E = \frac{mv^2}{2} + \frac{\alpha \mbox{e}^{\lambda}}{r} \, \, \, \, \, ; \, \, \, \, \,
\vec{J} = m \left( \vec{r} \times \vec{v} \right) + \alpha \hat{r} \, \, \, \, \, ; \, \, \, \, \,
L = |m \left( \vec{r} \times \vec{v} \right)|
\end{equation}

\noindent implying that the trajectories of the particle are confined to the same conical surface of the case of the
BPS monopole. The expressions for the coordinates $r(t)$ and $\phi(t)$ are the same as expressions
(\ref{sol:edonewton}), (\ref{sol:edonewton2})  and (\ref{phi}) as long as one replaces the constant $\alpha$ by the
"effective constant" $\overline{\alpha} = \alpha \mbox{e}^{\lambda}$. The conclusions are, once again, the same as in
Section \ref{sec:asympt} and bounded orbits occur only if $\overline{\alpha} < 0$ and $E < 0$.

Now, we study the first equation of (\ref{assintdyon}) in the ultra-relativistic limit. The expression obtained is:

\begin{equation}\label{ultrareldyon}
m\vec{a} = \frac{\alpha \sinh{\lambda}}{\gamma r^3} \left[ \vec{r} - \left( \vec{r} \cdot \vec{v} \right) \vec{v}
\right] + \frac{\alpha}{\gamma r^3} \left( \vec{r} \times \vec{v} \right)
\end{equation}

One constant of motion that can be found is the total energy of the particle:

\begin{equation}
E = m \gamma + \frac{\alpha \sinh{\lambda}}{r} = \mbox{cte}
\end{equation}

However, as we are considering the asymptotic ($r \gg 1$) and ultra-relativistic ($\gamma \gg 1$) limits, the first
term of the previous expression is very large compared to the second, and so we can take it as constant. Therefore, in
such case, the velocity of the particle is a constant of the motion, as in the situation of the BPS monopole.  It is
straightforward to see that the total relativistic angular momentum and the module of the orbital relativistic angular
momentum are also constants of motion, implying that the equations for $r(t)$ and $\phi (t)$ are the same as
(\ref{r_rel}) and (\ref{phi_rel}). Hence, the conclusions are the same as the ones in Section \ref{sec:asympt}: the
orbits are unbounded and restricted to the conical surface whose axis is $\vec{J}$ and whose half-opening angle is
given by (\ref{conerel}).

\section{Conclusions} \label{sec:concl}

It is possible to see that there is plenty of similarities between the two systems considered, i.e., a particle with
isospin interacting with a BPS monopole and with a Julia-Zee dyon. Both systems present no global equilibrium points
and allow the existence of radial motion consisting of unidimensional unbounded orbits. In this kind of motion,
although the isospin does not play a relevant role, remaining a constant of motion not coupled to the movement in the
Minkowskian spacetime, its sign is important to determine the shape of the potential energy and the location of the
possible return point, that indicates the direction in which the particle escapes.

Moreover, in the asymptotic region, considering the limits of low and high velocities, the two systems present orbits
confined to a conical surface whose axis lies in the direction of the total angular momentum. Although for an
intermediary velocity we can show that the orbits are not confined to a conical surface, we infer that they are
confined to the region given by the two distinct cones: the one referring to the Newtonian limit and the one referring
to the ultra-relativistic limit. For the same initial position of the particle, it is clear that the first cone has an
axis closer to the radial direction and an opening angle smaller than the cone related to the ultra-relativistic
limit. In addition, it is possible for both systems to allow stable closed orbits in the Newtonian limit, if the
energy of the particle is negative. In all of these asymptotic motions, the isospin is still not coupled to the
movement in the external space, but is no longer a constant of motion, as it precessionates in the internal space. The
radial component indeed seems to play the role of an "electric charge" in the Minkowskian spacetime and its sign is
fundamental in the shape of the effective potential energy, which determines the possibility of closed orbits.

What is interesting to notice is the analogy between this asymptotic motion and the motion of a single electric charge
in the presence of a magnetic monopole. As the fields of both systems, asymptotically, are interpreted as magnetic
monopoles, one could expect to recover at least some results from Classical Electromagnetism if the quadriforce and
isospin's evolution equations proposed were consistent. And we indeed obtained some of these results: in the limits of
high and low velocities, the particle is constrained to move over a conical surface, just as in the case of an
electric charge in the presence of a monopole (\cite{goldhaber}, \cite{schechter}). The radial
and azimuthal motions are not exactly the same in both cases, but this is not unexpected since we are dealing with two
very different systems in essence (e.g., the symmetry group of the systems involved in this work is $SU(2)$,
while for Electromagnetism, it is $U(1)$). Therefore, this simple analogy can give important clues about our choice
for the equation of the quadriforce and that of the isospin's precession.

Finally, we remark that, although the systems of a particle interacting with a BPS monopole and a particle interacting
with a Julia-Zee dyon seem very similar, they have important differences. In the special cases considered previously
(radial and asymptotic motion), these differences are revealed by the appearance of an extra factor multiplying the
radial component of the isospin. This is not unexpected, since the radial component is similar to an electric charge
and the new factor carries the constant of the dyon $\lambda$ that is responsible for the electric charge of the
entity.

\begin{acknowledgments}
The authors would like to acknowledge FAPESP and CNPQ for financial support.
\end{acknowledgments}

\end{document}